# Two-dimensional Dirac nodal-line semimetal protected by symmetry

Xingxia Cui[1†], Yafei Li[1†], Deping Guo[2†], Pengjie Guo[2,3,4†], Cancan Lou[1], Guangqiang Mei[1], Chen Lin[5], Shijing Tan[5], Zhengxin Liu[2], Kai Liu[2], Zhongyi Lu[2], Hrvoje Petek[6], Limin Cao[1], Wei Ji[2]* & Min Feng[1,7]*

[1]School of Physics and Technology and Key Laboratory of Artificial Micro- and Nano-Structures of Ministry of Education, Wuhan University, Wuhan 430072, China

[2]Beijing Key Laboratory of Optoelectronic Functional Materials & Micro-Nano Devices, Department of Physics, Renmin University of China, Beijing 100872, China

[3]Songshan Lake Materials Laboratory, Dongguan, Guangdong 523808, China

[4]Beijing National Laboratory for Condensed Matter Physics, and Institute of Physics, Chinese Academy of Sciences, Beijing 100190, China

[5]Hefei National Laboratory for Physical Sciences at the Microscale, University of Science and Technology of China, Hefei, Anhui 230026, China

[6]Department of Physics and Astronomy and Pittsburgh Quantum Institute, University of Pittsburgh, Pittsburgh, PA 15260, USA

[7]Institute for Advanced Studies, Wuhan University, Wuhan 430072, China

†These authors contributed equally to this study.
*Corresponding author. Email: wji@ruc.edu.cn (W.J.); fengmin@whu.edu.cn (M.F.)

**Dirac nodal line semimetals (DNLSs) host relativistic quasiparticles in their one-dimensional (1D) Dirac nodal line (DNL) bands that are protected by certain crystalline symmetries[1-3]. Their novel low-energy fermion quasiparticle excitations[4-7] and transport properties[8-10] invite studies of relativistic physics in the solid state where their linearly dispersing Dirac bands cross at continuous lines with four-fold degeneracy. In materials studied up to now, the four-fold degeneracy, however, has been vulnerable to suppression by the ubiquitous**

spin-orbit coupling (SOC)[11-14]. Despite the current effort to discover 3D DNLSs that are robust to SOC by theory[12,15], positive experimental evidence is yet to emerge. In 2D DNLSs, because of the decreased total density of states as compared with their 3D counterparts, it is anticipated that their physical properties would be dominated by the electronic states defined by the DNL. It has been even more challenging, however, to discover robust 2D DNLSs against SOC because of their lowered symmetry; no such materials have yet been predicted by theory. By combining molecular beam epitaxy growth, scanning tunneling microscopy (STM), non-contact atomic force microscopy (nc-AFM) characterisation, with density functional theory (DFT) calculations and space group theory analysis, here we reveal a novel class of 2D crystalline DNLSs that host the exact symmetry that protects them against SOC. The discovered quantum material is a brick phase tri-atomic layer Bi(110) [3-AL Bi(110)], whose symmetry protection and thermal stability are imparted by the compressive van der Waals (vdW) epitaxial growth on black phosphorus (BP) substrates. The BP substrate templates the growth of 3-AL Bi(110) nano-islands in a non-symmorphic space group structure. This crystalline symmetry protects the DNL electronic phase against SOC independent of any orbital or elemental factors. We theoretically[16] establish that this intrinsic symmetry imparts a general, robust protection of DNL in a series of isostructural 2D quantum materials. Combining the realization of this 2D structure through template growth and the generality of the intrinsic symmetry with robust DNL, our study provides the roadmap to explore the intrinsic 2D DNLS physics in other VA group elements with possible extension to a large class of real materials.

**Main**

Symmetry is the foundation for how atomic orbitals combine in crystalline environment to define the fundamental electronic properties of materials[17]. It describes the relativistic properties of charge carriers that have driven the discovery of novel electronic phases of the Dirac quantum materials[1,18]. Dirac nodal line semimetals (DNLSs) are a new class of quantum materials hosting relativistic quasiparticles with anticipated novel properties, such as enhanced correlation-driven physics[4,5], superconducting order parameters[7,8], anomalous magnetoresistance[10] and others[6,9]. These properties are anticipated from the robust fermionic excitations associated with four-fold degenerate Dirac nodes along 1D lines in the momentum space[1-3,19]. In 3D systems, it was proposed that the coexistence of spatial inversion (**I**) and time reversal (**T**) symmetry or mirror symmetry can protect the four-fold degenerate DNL without considering SOC[12-14,20,21]. However, SOC normally splits the four-fold DNL band into 2+2 pairs of degenerate bands[12-14,21]. Only in materials comprised with light-mass atoms, the SOC effect is insignificant and it was proposed that the four-fold degeneracy is approximately achieved[4,5,8,10,22-24]. Similar experimental and theoretical results suggest that DNLs in 2D materials also become gapped by SOC[25-29]. Symmetry may overcome this clash between DNL bands and SOC.

It was suggested that such protection of DNLs against SOC could be intrinsic to non-symmorphic symmetry where a fractional lattice translation may protect the Dirac point (DP) degeneracy[12,15,22,30]. A SOC gapping-resistant four-fold DNL state was recently predicted in a 3D system[12] with an **I** and **T** combined non-symmorphic operation. In some 2D materials, the non-symmorphic symmetry has been shown to protect the four-fold degeneracy of DPs at certain isolated k points against the SOC gapping[28]. However, no four-fold DNL state against SOC was found or even predicted in real 2D materials. Here, we describe, by experiment and theory, a brick phase three atomic-layer Bi(110) [3-AL Bi(110)] grown on a BP(110) crystal substrate. We show that this novel phase of Bi with the non-symmorphic $\mathcal{C}_{2v} \times Z_2^T$ symmetry is robust 2D DNLS that is protected against SOC. The existence of the DNL bands is experimentally established by imaging the quasi particle energy-band dispersion in low-temperature scanning tunneling spectroscopy (STS) measurements of coherent quasi particle interference (QPI). The established DNL bands are further verified by spatial conductance (d$I$/d$V$) mappings at the energies that the DNL states present. Our study reveals that the DNL bands near the Fermi level ($E_F$) of 3-AL Bi(110), where other trivial states are gapped, making it the "purest" DNLSs to be discovered and characterized. We further find that 3-AL Bi(110) features a non-trivial edge state coexisting with DNL, which is not predicted in non-symmorphic symmetry protected DNLS materials.

Bi nano-islands grow into a belt-like shapes exhibiting a BP-like Bi(110) structure onto a single-crystal BP(110) surface (Fig. 1a). The longer ZZ and shorter armchair (AC) directions of the islands nearly align with those of the anisotropic BP substrate, respectively (Extended Data Fig. 1). The well-known Bi(110) thin film usually grows in a bi-atomic layer (2-AL) BP-like structures[31-39] on various substrates. On those substrates, such bilayer Bi-films have measured STM heights of 6.60-7.00 Å[37-39]. The ~9.60 Å apparent height (Extended Data Fig. 2) of the Bi islands in our work, however, signifies a previously unknown form.

In STM images, each island appears as being composed of multiple ladder-like rails repeating with a ~50.0 Å period that extend for the entire island width (Fig. 1b), and have ~11.5 Å period rungs in the orthogonal (narrow) direction. The high-resolution STM image in Fig. 1c shows further contrast where three bright horizontal rungs composing an 8 $L_{ZZ}$ superlattice period, forming an 8 (ZZ) × 10 (AC) superlattice on BP (Fig. 1c). The subdivision of the 8 $L_{ZZ}$ contrast into three sub-periods of 3, 2.5 and 2.5 times $L_{ZZ}$ is evident in further expansion to atomic-resolution (Fig. 1d) and its associated contrast profile (Fig. 1e). To differentiate between the topographical and electronic origins to the ladder-like contrast, we perform a q-plus nc-AFM measurement. The sub-period contrast in the ZZ direction does not appear in the nc-AFM images (Fig.1f and 1g) consistent with its electronic origin most likely arising from the Moiré interference[31]. The nc-AFM image and its height profile along AC (Fig. 1f and 1g) show an appreciable atomic corrugation within the 10 $L_{AC}$ period.

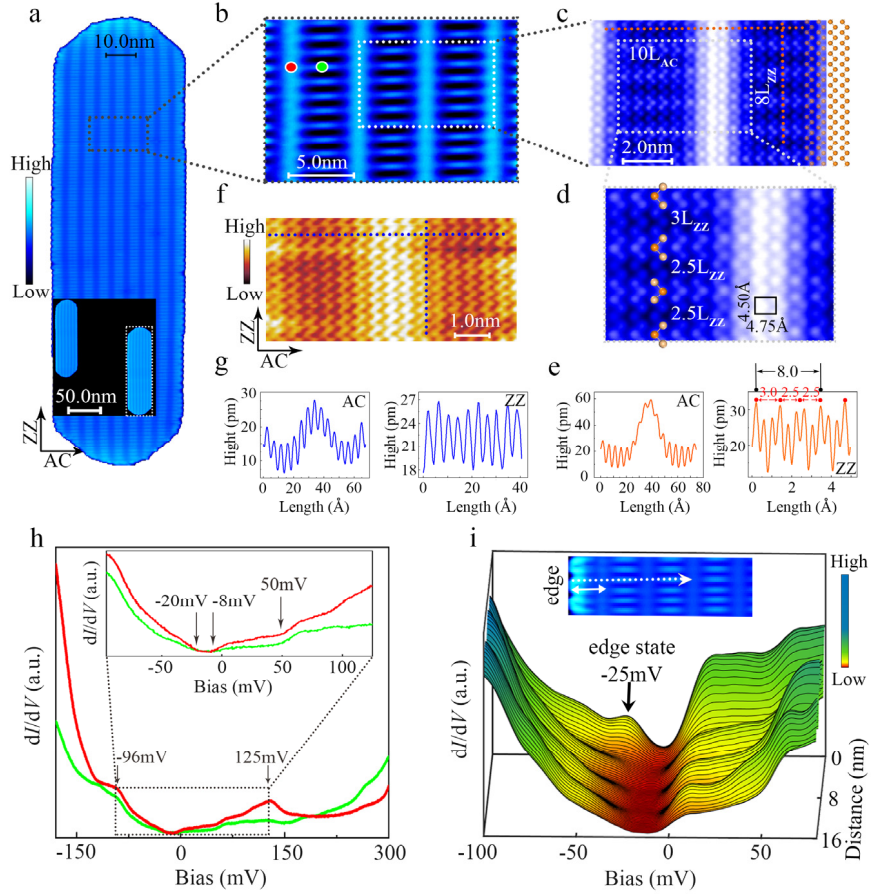

**Fig. 1 | Atomic structure and electronic spectra of the Bi(110) superlattice grown on BP(110) substrate. a,** STM image ($V_{bias}$ = +400 mV, $I_t$ = 100 pA) of a typical Bi island. The inset is a survey showing two islands. **b,** The highlight of the STM contrast inside the rectangle area in (a) showing the ladder contrast. **c,** Atom-resolved STM image ($V_{bias}$ = +100 mV, $I_t$ = 100 pA) of the rectangle area in (b). Atomic ZZ chains are clearly resolved. The dotted white rectangle represents the unit of the superlattice ladder with $8L_{ZZ} \times 10L_{AC}$. **d,** The enlarged image highlights the STM contrast within one unit of the superlattice. The orange balls represent the ZZ contrast of Bi atoms, identifying the three sub-periods in $8L_{ZZ}$. The black rectangle shows the lattice constants of the Bi(110) unit cell. **e,** The STM line profiles along the ZZ and AC directions, cutting at the dashed lines in (c). **f,** Q-plus AFM image of the superlattice. **g,** The AFM line profiles along the ZZ and AC directions, cutting at the dashed lines in (f). **h,** d$I$/d$V$ spectra acquired on two typical central locations of an island (dots in (b)). The inset shows amplified spectra over a small bias range. **i,** A series of d$I$/d$V$ spectra acquired along the dotted white arrow in the inset STM image.

The apparent height aside, our Bi islands also possess distinct electronic features from those of 2-AL Bi(110) thin-films[31-36]. Figure 1h presents typical d$I$/d$V$ STS spectra acquired at locations marked by the green and red dots in Fig. 1b, respectively. Those two spectra, despite originating from domains of different STM contrasts, show similar features and a consistently asymmetric V-shaped dip spanning from -96 mV to +125 mV. Such spectral feature portends to 2D Dirac states passing through a Dirac point[40-43]. However, the dip does not end up to an energy point, but to an energy plateau from -20 to -8mV (Fig. 1h inset). Another energy plateau, from -8 to +50mV, emerges after this flat dip. The results imply a different origin of the spectra with that for a 2D Dirac state. When the STM tip is scanned towards an edge of the island, a new peak, however, appears at approximately -25 mV (Fig. 1i). This feature is prominent at island edges and penetrates to a depth of ~38.0 Å (indicated by the solid arrow in Fig. 1i inset). The dip, the plateaus, the edge state, and the electronic sub-period contrast appear independent of the island width (Extended Data Fig. 3), which rules out their origin in quantum confinement. Thus, all these features differentiate our Bi(110) thin-films from either a topological insulator with nontrivial edge states[33,35] or a semimetal without edge state[36] in previously found Bi(110) thin-films.

The atomic structure and electronic spectroscopy point to our Bi islands exhibiting properties of a novel quantum allotrope, which is confirmed by our DFT calculations. Given the apparent height of ~9.6 Å, we consider four distinct tri-atomic layer (3-AL) Bi(110) structures (Extended Data Fig. 4); the most likely one is presented in perspective and side-views in Fig. 2a and 2b. Theory predicts this structure to be ~40 meV/Bi more stable than the other ones that, and its relative stability increases for thicker layers (Extended Data Fig. 4 and Extended Data Table 1), confirming its assignment. In this structure, the top and bottom Bi layers form two symmetrically disposed BP-like layers shared by a middle uncorrugated mono-atomic layer. Each top or bottom Bi atom bonds to three nearest neighbors and to four nearest neighbors for each middle layer Bi atom. Thus, the islands adopt a Bi(110)-like surface structure with a rectangular lattice of 4.53 times 4.85 Å. In a side-view, it appears as a wall tessellated by "bricks" consisting of pairs of Bi atoms, thus, we name it the "brick phase". Theory predicts that a freestanding brick phase 3-AL Bi(110) is prone to disproportionate into a hetero-layer of 2-AL and 4-AL. However, compressive strain applied through the van der Waals (vdW) epitaxy relation from the BP substrate is essential to stabilize the brick phase (Extended Data Table 1).

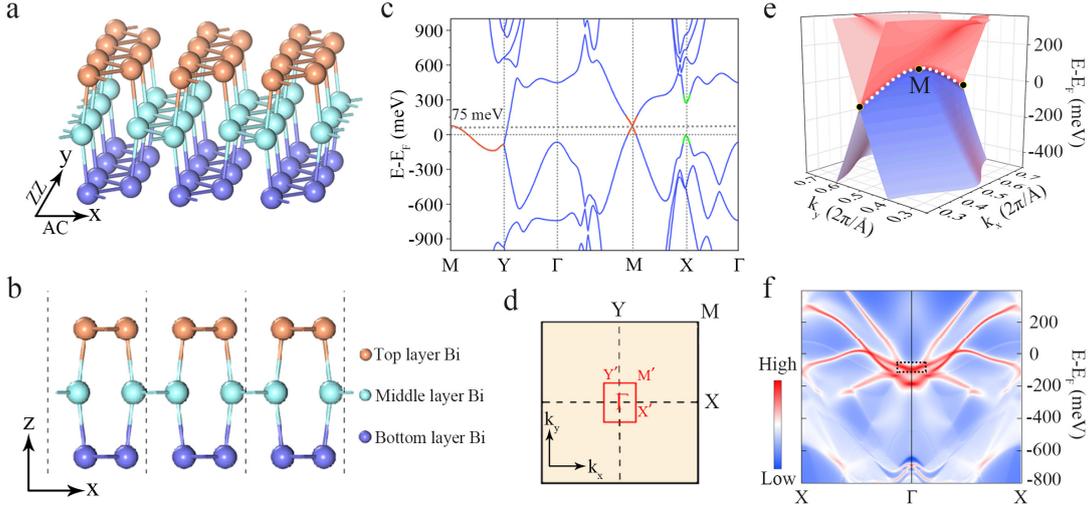

**Fig. 2 | Geometry and electronic band structure of the brick phase 3-AL Bi(110). a, b,** Top (a) and side (b) views of the atomic structure of the brick phase 3-AL Bi(110). **c,** DFT calculated electronic band structures of the brick phase 3-AL Bi(110) 2D infinite solid along high-symmetry directions with SOC. The Dirac cone at the M point (+75 meV) and the Dirac nodal line along Y-M are highlighted in red. The green curves emphasize the SOC gap opening at the X point. **d,** Corresponding schematic top view of the 1st BZ. The red coloured BZ represents the Bi(110) superlattice discussed later. **e,** Plot of a 3D electronic band surface of 3-AL Bi(110) with SOC with the dotted line tracing the DNL. **f,** Edge state of 3-AL Bi(110) along the AC direction with the dotted rectangle marking its energy.

Figure 2c shows the calculated electronic band structure with SOC of a freestanding brick phase 3-AL Bi(110) along the paths connecting the high symmetry points illustrated in Fig. 2d; it shows a semimetal with a gap opening at a time-reversal invariant point X (indicated by the green curves). Within the band-gap, the highest valence and the lowest conduction bands contact at the M point (+75 meV; the red cross), which forms a four-fold degenerate nodal point with, as proved by our theory[16], linear dispersion in the $k_y$ direction (Fig. 2e). Such linear dispersion indicates that the nodal point is a Dirac point, which at the boundary of the Brillouin zone (BZ) extends in the Y-M line along the $k_x$ direction to form a DNL (the red curve). Figure 2e shows a 3D presentation of the DNL, illustrating how the nodal line connects the four-fold degenerate Dirac bands in the Y-M line that disperse linearly orthogonal to the nodal line.

The inclusion of SOC only flattens the dispersion of the DNL rather than reducing its degeneracy to 2+2, suggesting that the DNL character in 3-AL Bi(110) is protected against SOC[16]. The bandgap at the X point is opened by SOC induced band inversion[16], which offers nontrivial edge states at the boundary of the Bi thin films. As shown in Fig. 2f, a flat band just below $E_F$, marked in the black dashed rectangle, was found in the Γ–X direction. This most likely explains the experimentally observed edge state at -25 meV (Fig. 1i). Therefore, the brick phase 3-AL Bi(110) has nontrivial edge states that coexist with the DNL.

Further DFT calculations robustly reproduce the DNL bands in the grown $8L_{ZZ} \times 10L_{AC}$ Bi(110)/BP superlattice. Figure 3a shows the top and side views of the fully relaxed atomic structure used to model the supported 3-AL Bi(110) superlattice; it shows that the three sub-periods in the ZZ direction originate from the lattice mismatch between Bi and BP (Fig. 3b), causing the characteristic texturing of the STM contrast (Fig. 1d), which is reproduced by STM simulation (Fig. 3c). A regular vertical corrugation of 8-16 pm is found along the AC direction, consistent with the AFM results (Fig. 1f). The agreement between the experiment and theory confirms the assignment of proposed structure.

Considering the 3-AL $8L_{ZZ} \times 10L_{AC}$ Bi(110)/BP superlattice, the 1×1 unit-cell BZ of freestanding 3-AL Bi(110) folds to an eight- by ten-times smaller superlattice BZ (SLBZ, Fig. 2d) with its boundaries designated by X´, Y´, and M´. In the SLBZ the original Y-M line, that supports the DNL, folds to the Γ–X´ line (because the band folds nine times along the $k_y$ direction, the original Y point becomes folded into the Γ point in the SLBZ). The calculated band structure of the superlattice is substantially folded such that the Bragg reflection of DNL produces rhombus-like shapes in the band structure plotted in Fig. 3d and 3e. In Fig. 3d those red triangles highlight the rhombus formed by the pure DNL after folding. In Fig. 3e, the red dots highlight the DNL which arises through hybridization with the $p_z$ orbitals of P atoms underneath at 17 meV to ~152 meV. Consequently, the hybridization modestly lifts the fourfold DNL degeneracy by causing a ~6 meV splitting.

A finite value plateau appeared in DOS within the energy range of the DNL

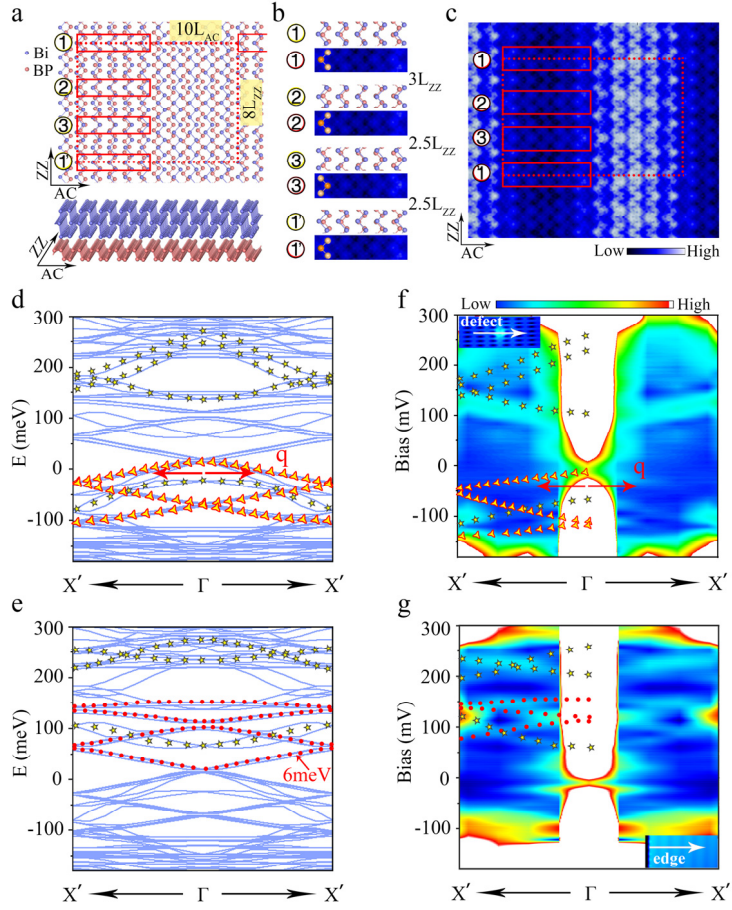

**Fig. 3 Atomic model and E(q) dispersions of the folded Dirac nodal lines in 3-AL Bi(110) superlattice on BP surface. a,** Calculated atomic model of the $8L_{ZZ}\times10L_{AC}$ Bi (110) superlattice on BP surface. The red dotted-rectangle defines the superlattice edge. The solid red rectangles highlight the atomic registration of the Bi atoms on BP surface, which causes the three sub-periods within the $8L_{ZZ}$ period. The enlarged images are given in (b). **c,** Simulated STM image of the superlattice ($V_{bias}$=+100 mV) with the solid red rectangles marking the apparent STM contrast of the regions corresponding to those in (a). The amplified STM contrast images are placed in (b) to show the STM contrast orientation (marked by the orange balls) of the Bi ZZ chain. **d, e,** Calculated band structures in SLBZ along the X´–Γ–X´ direction of the superlattice highlighting the folded pure DNL state using red triangles (d) and the folded hybridized DNL state using red dots (e). The green-lined stars represent the electronic bands not related to the DNL state but observable in the experiments. The same scheme of markers is applied to (f) and (g). **f, g,** Measured E(q) dispersions in SLBZ along X´–Γ–X´ obtained from QPI scattering by a surface defect (f) or an island edge (g). In the QPI patterns, the wave vector $q$ represents allowed scattering of the Bloch electron waves from the initial state to the final state.

state is a characteristic feature of an ideal linearly dispersed DNL in 2D DNLS, different from the V-dip of 3D DNLSs. In the present superlattice case, the BP-Bi hybridized states develop from ~-90 to +100mV with a gap from 17 to 60 mV, as highlighted by the green-lined stars in Fig. 3d and 3e. This band structure partially explains the observed flat dip from -20 to -8 mV and the plateau from -8 to 50 meV in the d$I$/d$V$ spectra (Fig. 1h). The DNL states were also verified by our d$I$/d$V$ mapping of a serious of energies within the DNL state where we did experimentally observe nearly identical spatial LDOS distribution patterns at those DNL state energies (from -90mV to 128mV) (Extended Data Fig. 5). These results indicate the 2D DNL band origin of the plateau-like LDOS.

The superlattice folding and the substrate's interference complicate the DNL appearance in the *k*-space, which obscures our DNL characterization by methods such as angle-resolved photoemission spectroscopy. We thus employ the real-space method of the STS quasi particle interference (STS-QPI) to extract the energy-momentum [E(q)] dispersion of the bulk and surface bands[44,45].

Our QPI spectra, again, confirm the theoretically predicted DNL that is modified by the superlattice. Figure 3f and 3g present the QPI imaging of the E(q) dispersion along the X′–Γ–X′ line (see Extended Data Fig. 6 for more details). In Fig. 3f, red triangles are overlaid to guide the eye to linearly dispersing bands, appearing as the theoretically predicted rhombus structure of the pure DNL band (Red triangle in Fig. 3d). Stronger scattering at the island edges (Fig. 3g) sharpens the QPI contrast acquired there, especially for those hybridized DNL bands. In this case, the rhombically folded hybridized DNL band dispersions are observed from 90 to 155 mV (red dots in Fig. 3g), consistent with the theoretically predicted energy window from 17 to 152 meV (red dots in Fig. 3e). Similar measurements revealing the E(q) relations along Y′–Γ–Y′ (the ZZ direction, Extended Data Fig. 7) further confirm our interpretation of STS-QPI measurements. These results are complementary and reproducible, which unequivocally confirm the direct observation of DNLs in 3-AL Bi(110).

Space group theory analysis[16] elucidates how the crystal symmetry of 3-AL Bi(110) protects its DNL character. It belongs to a non-symmorphic space group

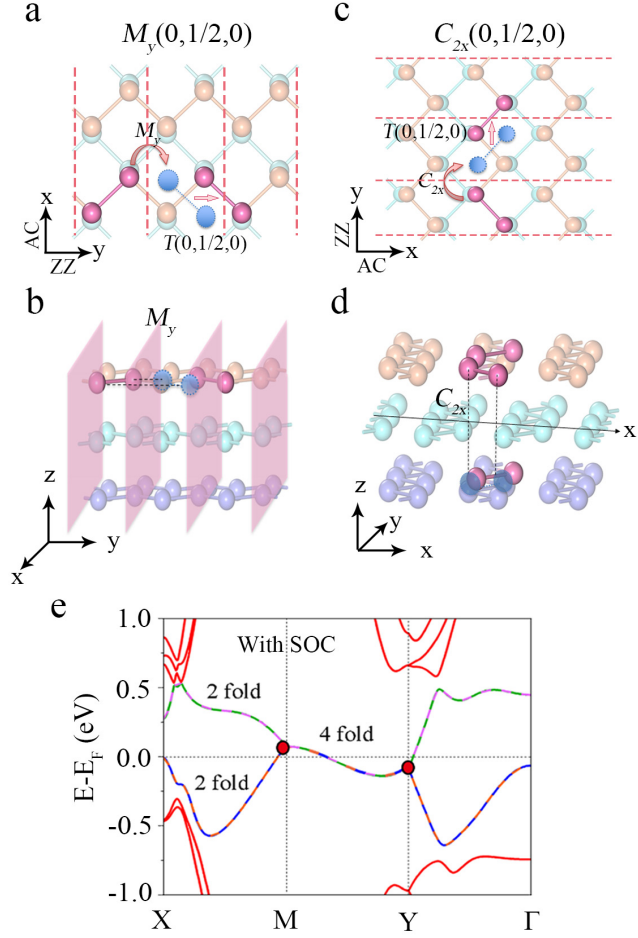

**Fig. 4 | Space group symmetry analysis of the brick phase 3-AL Bi (110). a, b,** The atomic model shows operations of {$C_{2x}$| (0, 1/2, 0)} from the top (a) and side (b)views of 3-AL Bi(110). The pink atoms represent the initial and final positions of Bi atoms involved in the operation and the light-blue atoms represent the transition positions of the atoms before taking the fractional lattice transition. **c, d,** Atomic model shows operations of {$M_y$| (0, 1/2, 0)} from the top (c) and side (d) views of 3-AL Bi(110). **e,** Band structures of 3-AL Bi(110) appearing along Y-M SOC. The curves of mixed pink and blue colors represent the origin of the fourfold degenerate DNL bands.

Pmma, whose point group G = $D_{2h} \times Z_2^T$ is generated by the $C_{2x}$, $C_{2y}$, $M_z$ and $\tilde{T}$ symmetry operations. Here $\tilde{T}$=**IT** is a sequence of inversion and translation symmetry operations. It, in combination with $\tilde{T}^2$=-1, guarantees a two-fold Kramers degeneracy in the spin degree of freedom at any *k* point, which is necessary to obtain a DNL against SOC. Such spin degeneracy is also reflected by the anticommutation relation between two non-symmorphic symmetry operations from group $C_{2v} \times Z_2^T$, namely {$M_y$| (0, 1/2, 0)} (Fig. 4a and 4b) and {$C_{2x}$| (0, 1/2, 0)} (Fig. 4c and 4d) on the lattice. Another anticommutation relation between the {$M_y$| (0, 1/2, 0)} and **IT** operations[16], resulting in an extra two-fold orbital degeneracy along the Y-M ($k_x$, π, 0) line under SOC. Thus, the four-fold degeneracy is resistant to strong SOC (Fig. 4e). We conclude, therefore, that the non-symmorphic symmetry plays the essential role of protecting the quadruple degeneracy against SOC in this 3-AL brick phase Bi(110).

In other words, strong SOC does not, as a usual consequence of it, lifts the degeneracy of the DNL in this novel 3-AL brick phase Bi(110) layer, but rather introduces some unique electronic features. Particularly, the SOC produces a smaller dispersion of the DNL band around $E_F$, which together with the band gap it opens around $E_F$, stands out the DNL from diminishing its character by mixing with other electronic states. These two hallmarks facilitate future probe and manipulation of the DNL fermions within the SOC gap. In addition, the strong SOC leads to a band inversion that engenders a non-trivial edge state on the 3-AL Bi(110) surface, enabling coexistence of non-symmorphic symmetry protected DNL and topological non-trivial edge state in the same materials.

We emphasize that the DNL in brick phase 3-AL Bi(110) is dictated by the crystal symmetry. This discovered structural symmetry is, to the best of our knowledge, the exact one in 2D that, intrinsically guarantees the existence of quadruple-degenerate DNL resistance to SOC[16]. The DNL is, therefore, promised to exist in any system that satisfies the above symmetry, e.g. the brick phase 5-AL Bi, 3-AL Sb, As or P. Our DFT calculations verify that 1- and 5-AL Bi(110), as well as 3-AL P(110), As(110) and Sb(110) all form DNLS solids that are protected against SOC[16]. Experimentally, we confirm the symmetry protection by observing similar STM ladder contrast and robust edge states in Bi islands that we judge to have 5-AL Bi(110) structure (Extended Data Fig. 8). We also expect 5-AL or thicker Bi(110) thin films to host DNLs that are less influenced by the substrate than the 3-AL Bi(110) islands. We realized that the BP substrate acts as a critical template for the unique brick phase 3-AL Bi(110) growth. The anisotropic interlayer vdW interactions at the Bi–BP interfaces enable the in–plane confinement that guides and stabilizes the growth of the brick phase structure. We expect that the vdW growth strategy has potential to enable similar brick phase materials to be grown with most VA group elements and related materials, to expand the experimental study of intrinsic 2D DNLS quasiparticle physics in real materials.

**Methods**

**Sample preparation and STM/STS measurements.** The BP crystals are self-grown by chemical vapor transport (CVT) method using red phosphorus, tin iodide ($SnI_4$), and tin powders as starting materials, in a two-zone tube furnace with a temperature gradient of 600°C to 540°C. The STM and spectroscopy experiments are carried out in an ultrahigh vacuum low temperature STM system (CreaTec). Prior to STM experiments, the BP crystals are cleaved *in-situ* in a preparation chamber under ultrahigh vacuum at room temperature (RT). Bismuth atoms (99.999% purity, Aldrich) are evaporated from a resistively heated evaporator onto a freshly prepared BP surface. The BP substrates are kept at RT during the evaporation. The prepared sample is then immediately transferred into the STM chamber, and cooled down to 77 K and/or 4.5 K. STM topographic images are acquired in the constant-current mode. The d$I$/d$V$ spectra are measured using the standard lock-in technique with a bias modulation of 8 mV at 321.333 Hz. The STM tips are chemically etched tungsten, which are further calibrated spectroscopically against the Shockley surface states of cleaned Cu(111) or Au(111) surfaces before performing measurements on Bi islands/BP. The STS-QPI technique characterizes the surface band structure near $E_f$ through recording surface standing waves induced by electron scattering in proximity of point defects (or edges). The E(q) dispersion is obtained by a fast Fourier transform (FFT) analysis of the spatial tunneling conductance images from a series of differential conductance d$I$/d$V$ spectra.

The quasiparticle interference signal is acquired as a function of distance over 150 Å from a defect along the *x* (AC) direction.

**DFT calculations.** DFT calculations are performed using the generalized gradient approximation for the exchange-correlation potential, a plane-wave basis, and the projector augmented wave method, as implemented in the Vienna *ab-initio* simulation package (VASP)[46-48]. The pure Bi films are modeled by utilizing a 2D freestanding three atomic layer Bi quantum structures separated by 15 Å vacuum layers. The Bi-BP superstructure is modeled using a superlattice consisting of a 8$L_{ZZ}$ ×10$L_{AC}$ 3-AL Bi(110) and 11$L_{ZZ}$ ×11$L_{AC}$ 2-AL BP(110) with a 15 Å vacuum layer perpendicular to the Bi and BP layers. The energy cutoff for plane wave is set to 650 and 350 eV for variable volume structural relaxation of pure Bi 2D materials and invariant volume structural relaxation of Bi island on BP surface. For freestanding Bi layers, the k-points sampling of the first Brillouin zone is 14×14×1, generated automatically by Monkhorst-Pack method[49]. The energy cutoff for plane wave is set to 650 and 350 eV for variable volume structural relaxation of pure Bi 2D materials and invariant volume structural relaxation of Bi-BP superstructure. The k-points sampling includes only the Γ-point for Bi-BP superstructure. The bottom BP layer is kept fixed and all other atoms are fully relaxed until the residual force per atom is less than 0.05 eV/Å during the relaxations of Bi-BP superstructure and less than 0.001 eV/Å during all the freestanding pure Bi relaxations. In structural relaxation and electronic property calculations, DFT-D3 dispersion correction method is used with the Perdew-Burke-Ernzerhof (PBE) exchange functional (PBE-D3)[50,51]. All electronic properties of the Bi-BP superstructure are calculated with the consideration of spin-orbit coupling (SOC) by setting LSORBIT = .TRUE. in VASP software. The STM image simulation is performed using the Tersoff-Hamann method[52]. The edge states are obtained by creating a 1D 3-ALBi nanoribbon and by calculating the electronic states populating the spatial locations into the vacuum by using Wannier90[53] and WannierTools[54]. The 5d, 6s and 6p orbitals of Bi are used to construct the maximally localized Wannier functions. The spectrum of edge states is calculated by the iterative Green's function using the tight-binding Hamiltonian generated by Wannier90 and illustrated by gnuplot.

**Data availability:** The data that support the findings of this study are available from the corresponding authors upon request.

**Acknowledgement**

We gratefully acknowledge fruitful discussions with Haiwen Liu, Hongjian Du, Shaochun Li, Rui Yu, Chunwei Lin, Hongming Weng and Xi Dai. This work is supported by the National Key R&D Program of China (Grant Nos. 2017YFA0303500, 2017YFA0303504, 2018YFE0202700); the Strategic Priority Research Program of Chinese Academy of Sciences (Grant No. XD30000000), the National Natural Science Foundation of China (Grant Nos. 11774267, 61674171 and 11974422). The Fundamental Research Funds for the Central Universities and the Research Funds of Renmin University of China [Grants No. 16XNLQ01 and No. 19XNQ025 (W.J.) and No. 21XNH090 (D.P.G.)]. P.J.G. was supported by a China Postdoctoral Science Foundation–funded project (GrantNo.2020TQ0347). The contributions from H.P. thank the Luo Jia Visiting Chair Professorship in Wuhan University. Calculations were performed at the Physics Lab of High-Performance Computing of Renmin University of China and Shanghai Supercomputer Center.

**Author Contributions**

X.X.C, Y.F.L and C.C.L grew the Bi samples and performed STM measurements; D.P.G. and W.J. performed first principle calculations; P.J.G., Z.X.L and W.J. conducted the symmetry analysis; G.Q. Mei and L.M.C. grew the BP crystals; M.F. and L.M.C initiated the project and experiments; W. J. conceived the theoretical calculations and analysis; C. L. and S. J. T. participated in experiments of electronic structure studies; K.L. and Z.Y.L. participated in data analysis and discussions; M.F., W.J., H.P. and L.M.C. analyzed the data, and wrote the manuscript with input from all authors.

**Additional information**

Additional data related to this paper are available from the corresponding authors upon request.

**Competing interests:** The authors declare that they have no competing interests.

# Extended Data

# Two-dimensional Dirac nodal-line semimetal protected by symmetry

Xingxia Cui[1†], Yafei Li[1†], Deping Guo[2†], Pengjie Guo[2,3,4†], Cancan Lou[1], Guangqiang Mei[1], Chen Lin[5], Shijing Tan[5], Zhengxin Liu[2], Kai Liu[2], Zhongyi Lu[2], Hrvoje Petek[6], Limin Cao[1], Wei Ji[2]* & Min Feng[1,7]*

[1]School of Physics and Technology and Key Laboratory of Artificial Micro- and Nano-Structures of Ministry of Education, Wuhan University, Wuhan 430072, China

[2]Beijing Key Laboratory of Optoelectronic Functional Materials & Micro-Nano Devices, Department of Physics, Renmin University of China, Beijing 100872, China

[3]Songshan Lake Materials Laboratory, Dongguan, Guangdong 523808, China

[4]Beijing National Laboratory for Condensed Matter Physics, and Institute of Physics, Chinese Academy of Sciences, Beijing 100190, China

[5]Hefei National Laboratory for Physical Sciences at the Microscale, University of Science and Technology of China, Hefei, Anhui 230026, China

[6]Department of Physics and Astronomy and Pittsburgh Quantum Institute, University of Pittsburgh, Pittsburgh, PA 15260, USA

[7]Institute for Advanced Studies, Wuhan University, Wuhan 430072, China

†These authors contributed equally to this work.

*Corresponding author. Email: wji@ruc.edu.cn (W.J.); fengmin@whu.edu.cn (M.F.)

**Extended Data Figure 1**

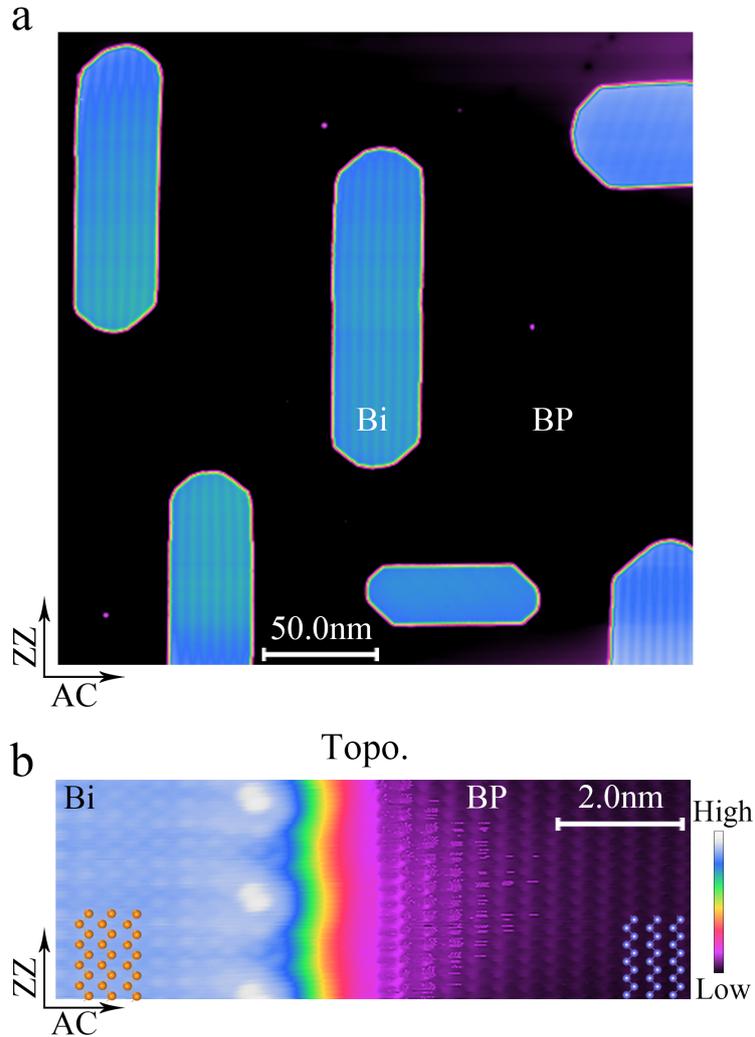

**Extended Fig. 1 STM topographic images of Bi(110) islands grown on BP surface**. **a,** Large scale STM image showing that the islands grow with a preferred azimuth along the zigzag (ZZ) direction of the substrate. **b,** High resolution STM topographic height image showing the atomic contrast within a Bi island and the BP substrate. Despite the big height difference, it is clear that the BP zigzag chains and Bi zigzag chains are oriented almost parallel. The images are obtained at LN$_2$ temperature.

**Extended Data Figure 2**

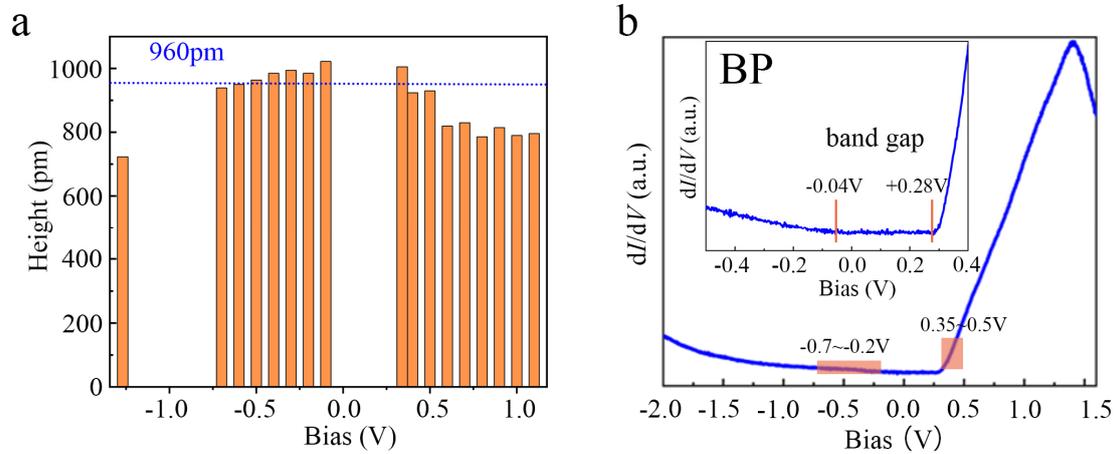

**Extended Fig. 2 STM measured height of Bi(110) islands on BP surface for different STM bias conditions. a,** The height of the Bi islands varies upon different bias and centers at ~960 pm in the bias range from -0.7 V to -0.2 V and 0.35 V to 0.5 V . We suggest this variation occurs because at the bias outside from this energy range, the BP substrate has a higher density of states (DOS), as shown in **b**. This makes the STM measured height of Bi islands on BP smaller than the actual value. This phenomenon is more obvious at positive voltages, where the BP substrate hosts a very high DOS in the conduction band (Guo et al., Linear Scanning Tunneling Spectroscopy Over A Large Energy Range in Black Phosphorus, J. Appl. Phys. 124, 045301 (2018)). To exclude the influence of the DOS from BP substrate, we conclude that the measured height under the bias around the band gap edges is close to the actual value. The measured average 960 pm height in this energy range is consistent with three atomic layer height of Bi(110) crystalline structure.

**Extended Data Figure 3**

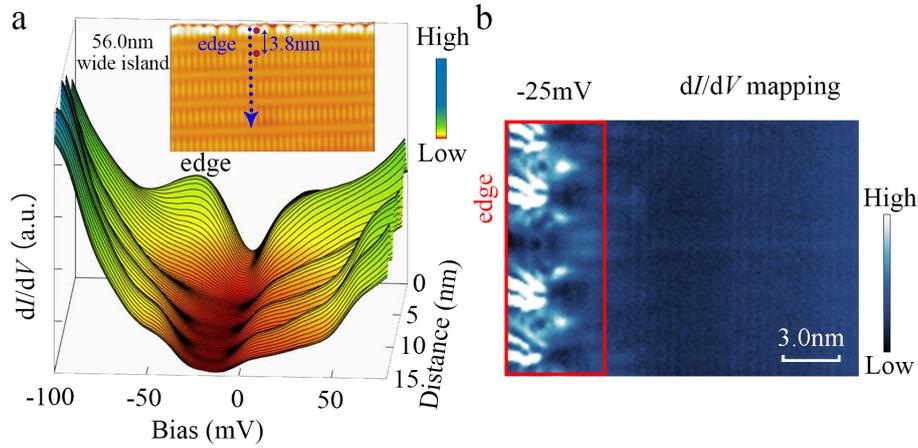

**Extended Data Fig. 3 Topological edge states of Bi(110) superlattice grown on BP(110) substrate**. **a,** Series STS d$I$/d$V$ spectra measured along a line normal to Bi edge in the zigzag direction for a 56 nm wide island. The island shows an edge state at -25 mV, which is the same as for the 33 nm wide island in Fig. 1i. The edge state penetrates into the island for ~3.8 nm, similar to that observed in Fig. 1i. **b,** d$I$/d$V$ mapping acquired at -25 mV. The bright contrast at the edge denotes the real space location of the edge state. Note that the edge state has structured contrast. This uneven contrast might arise from a partly irregular atomic organization at edges and the hybridization of edge state with substrate (Drozdov et al., One-dimensional topological edge states of bismuth bilayers, Nat. Phys. 10, 664-669 (2014)). The robustness and reproducibility of the measured edge state in islands with different sizes eliminate the possibility of quantum confinement causing a modulation in the measured DOS.

**Extended Data Figure 4**

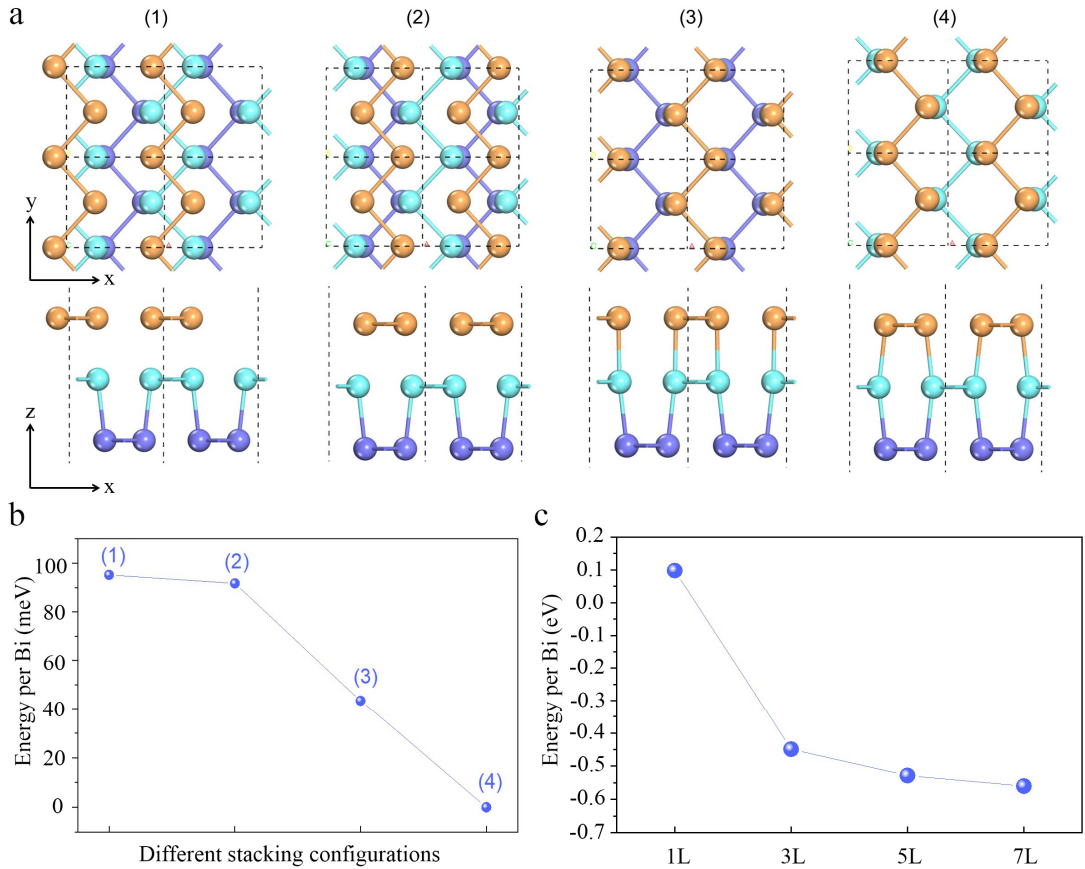

**Extend Data Fig. 4 The calculated Bi(110) tri-layers with different stacking configurations and energies. a,** Four different stacking configurations (1-4) considered of Bi tri-layers (atoms belonging to different layers are presented in different colors). **b,** Energy per Bi atom of the four different stacking configurations of Bi tri-layers. Configuration (4) has ~40 meV/Bi atom lower energy than that the next higher configuration (3), showing that it is the most stable one. **c,** The total energy of different odd Bi(110) layers showing that the relative stability increases for thicker layers.

**Extended Data Table 1**

| Layer | Total energy (meV/Bi) | |
|---|---|---|
| | freestanding lattice (a= 4.85 Å, b= 4.53 Å) | constrained lattice on BP (a= 4.70 Å, b= 4.50 Å) |
| 2-AL | 104.1 | 123.2 |
| 3-AL | 59.2 | 58.4 |
| 4-AL | 0 | 0 |
| (2-AL+4-AL)/2 | 52.05 | 59.52 |
| ΔE | 7.2 | -3.2 |

**Extended Data Table 1 Total energy of different freestanding Bi atomic layers and on BP substrate.** The total energy of four-atomic (4-AL) layer Bi was set to zero. The total energy of freestanding bi-atomic layer (2-AL), tri-atomic layer (3-AL) and 4-AL Bi is 104.1, 59.2 and 0 meV/Bi atom, respectively. $\Delta E = E_{3-AL} - (E_{2-AL} + E_{4-AL})/2$. Comparing the total energy of the (2-AL+4-AL)/2 and 3-AL Bi, it is found that the former energy is 7.2 meV/Bi lower than later. This indicates that 3-AL Bi is less stable than (2-AL+4-AL)/2 Bi without substrate. With the experimental lattice on BP substrate, the total energy of 2-AL, 3-AL and 4-AL Bi is 119, 58.3 and 0 meV/Bi atom, respectively. Comparing the total energy of the (2-AL+4-AL)/2 and 3-AL Bi, it is found that 3-AL Bi is 3.2 meV/Bi lower. This suggests that the 3-AL Bi structure is more stable when adsorbed on BP substrate.

**Extended Data Figure 5**

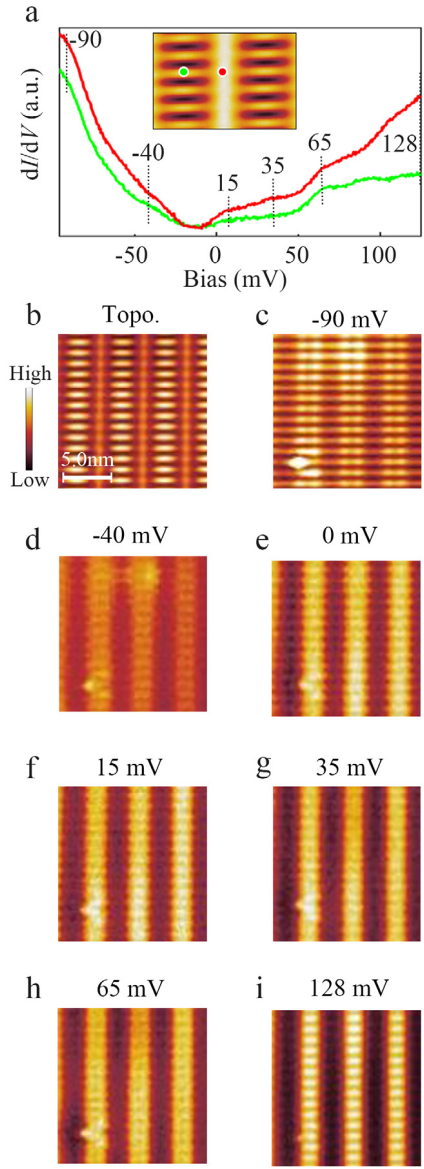

**Extended Fig. 5 STM d$I$/d$V$ spectra and mappings acquired on 3-AL Bi(110) superlattice on BP within the energy range of the DNL bands. a,** The d$I$/d$V$ spectra marked with the energy locations for the d$I$/d$V$ mapping measurement. **b** to **i,** d$I$/d$V$ mapping images taken at various bias voltages corresponding to the energy values of the around E$_f$ in (a). The mapping images are acquired on the areas far away from the long edges to eliminate the contributions from the edge scattering. Despite the atomic corrugation induced contrast, the real space distribution of the LDOS shows similar strip-like features at all the measured energies.

**Extended Data Figure 6**

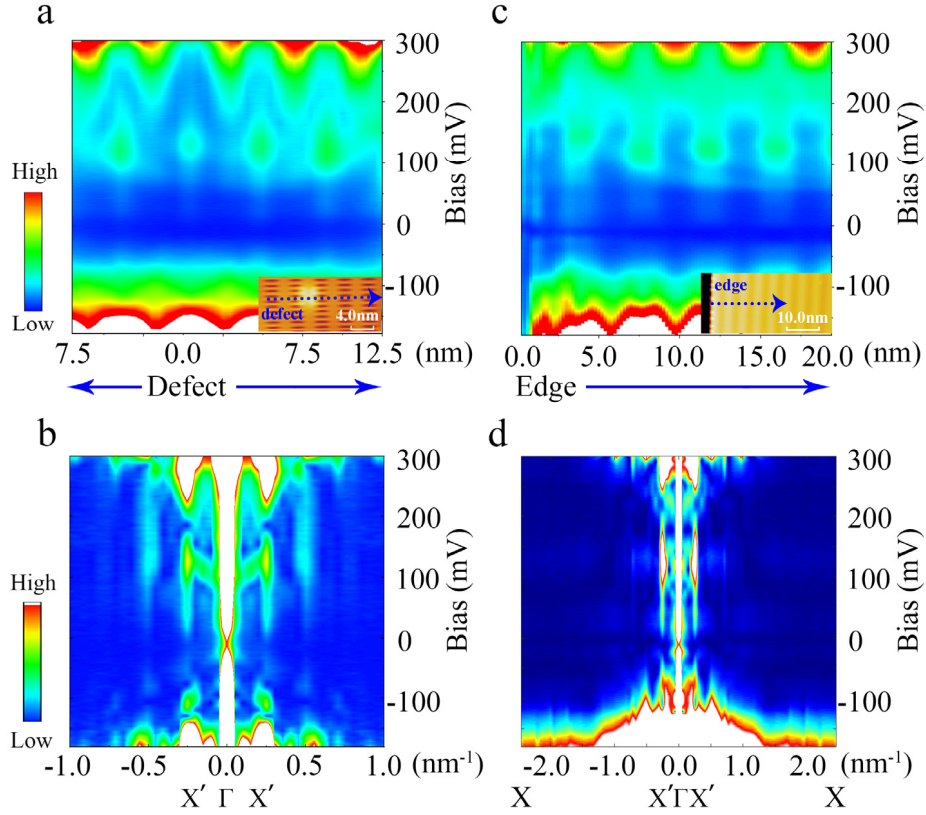

**Extended Data Fig. 6 2D d*I*/d*V* spectral map of series of d*I*/d*V* spectra along the AC direction (Γ-X) and the E(q) dispersions of 3-AL Bi(110) superlattice on BP surface.**
**a, c** 2D d*I*/d*V* spectral maps acquired along a line in the AC direction, by crossing a typical defect on a Bi surface (a) and from a Bi island edge (c). The insets in both images show locations where the spectra are acquired. One can detect the QPI features when approaching the defect and edge. **b, d** The obtained E(q) relations corresponding to (a) and (c) respectively by performing FFT of the corresponding series d*I*/d*V* spectra. In the E(q) images, X point is on the boundary of the 1st BZ corresponding to those the free-standing 3-AL Bi(110) (Fig. 2d). Because of the band folding by the superlattice, X′ point is on the boundary of the 1st SLBZ (Fig. 2d). Zooming into the momentum space range of Γ– X′ along Γ–X direction approaches the E(q) relations of the 1st SLBZ (Fig. 3f, Fig. 3g), corresponding to the QPI behavior of the grown Bi(110) superlattice. The white regions near the Γ point in the E(q) images result from a high FFT amplitude resulting from the structural atomic corrugation modulating STS spectra in the AC direction (Fig1). The periodic atomic corrugations also cause interference in scattering of the electrons from the edge or the defect, making the extracted E(q) dispersion very sensitive to the choice of the surface perturbation.

**Extended Data Figure 7**

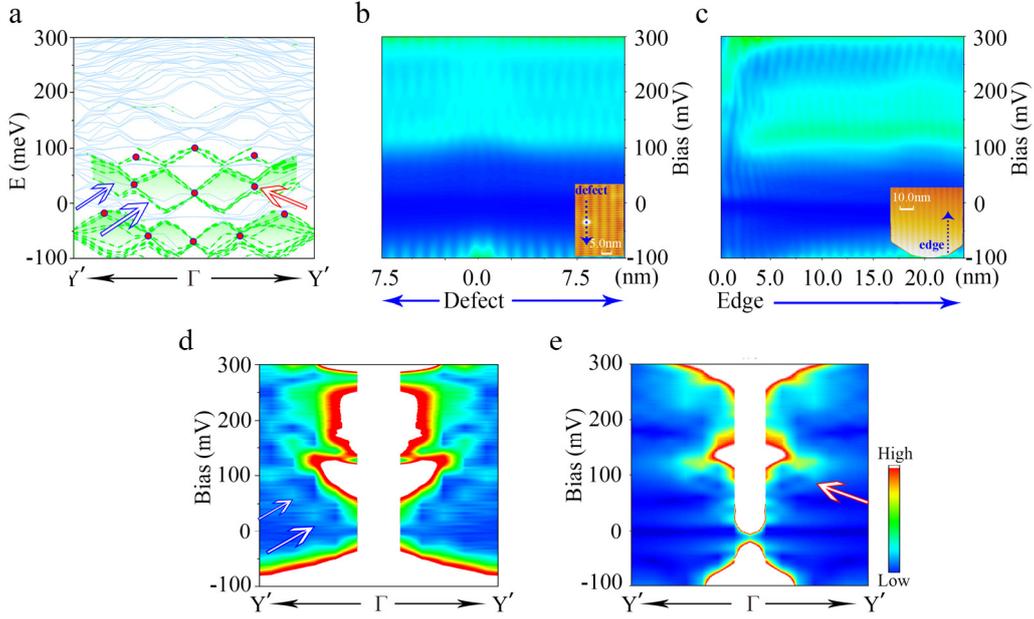

**Extended Data Fig. 7 Band E(q) dispersions of 3-AL Bi(110) superlattice on BP surface along the ZZ (Γ–Y) direction. a,** Calculated band structures along Y′–Γ–Y′ direction for Bi (110) $8L_{ZZ} \times 10L_{AC}$ superlattice on BP. The 1st BZ of the superlattice is shown as the inner red rectangle in Fig. 2d. The green color highlights the bowtie-shaped rhombi in the band structure along the Y′–Γ–Y′ direction, which is caused folding of linear bands from the X-M to the Γ–Y′ direction. The red dots highlight the folded Dirac cones at the Y point. **b, c,** 2D d$I$/d$V$ spectral map of a series d$I$/d$V$ spectra acquired on a line along the ZZ direction, crossing a typical defect on the surface (b) and from an edge (c) of the island. The insets in both images show the locations where the spectra are acquired. One can see the spectral changes caused by QPI when closing the defect and edge. **d, e,** E(q) dispersions along Y′–Γ–Y′ obtained from QPI scattering by a surface defect (d) and an edge (e) by performing FFT of the corresponding series of d$I$/d$V$ spectra. Compared with the experimental d$I$/d$V$ spectra, the DFT calculations elevate all of the state energies by ~28 meV. Several bow-tie shapes (marked by the blue-white arrows), originating from the crossing of the linear bands, are seen in (d). Because of the strong scattering at the edge, the QPI pattern is relatively sharp in (e) and a clear Dirac point appears in the QPI as

marked by the red-white arrow. The colored arrows in (d) and (e) point out the features of E(q) dispersions that are predicted in (a).

**Extended Data Figure 8**

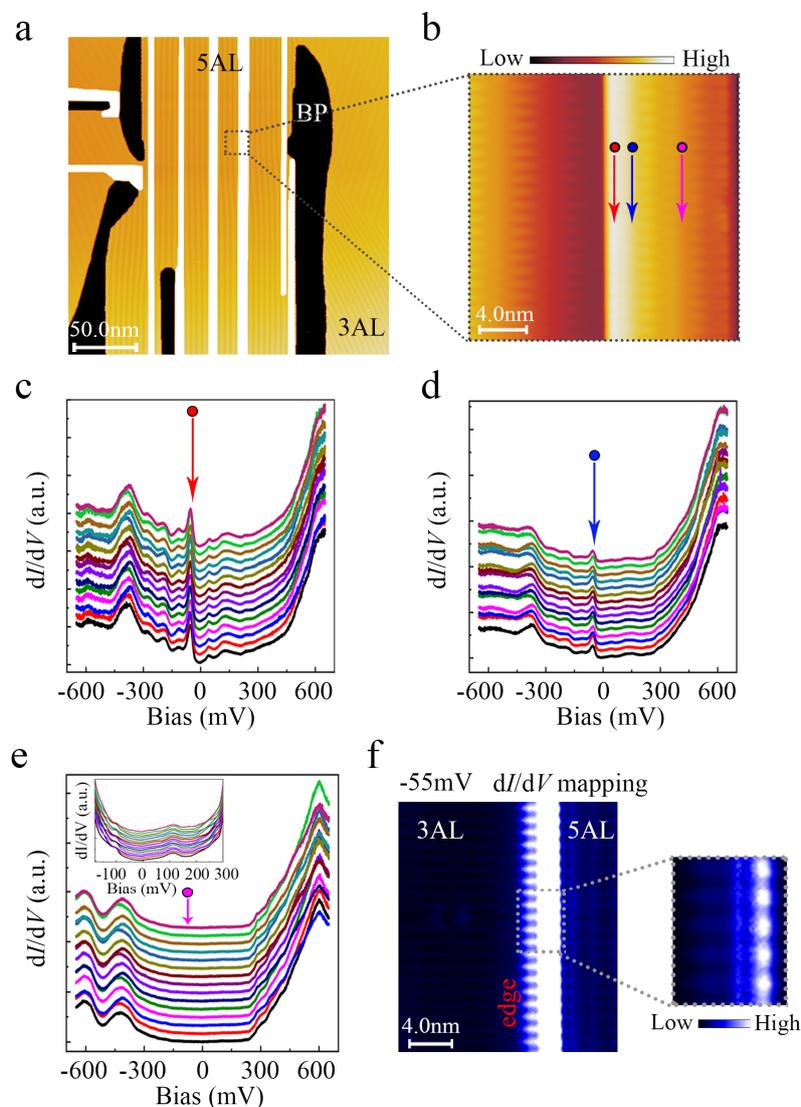

**Extended Data Fig. 8 Topographic STM images of 5-AL Bi(110) nanoribbons grown on BP substrates and the topological edge states measured by STS d*I*/d*V* spectra**. **a,** STM image shows the coexistence of 3-AL Bi and 5-AL Bi on BP substrate at a higher Bi coverage. When continuously increasing the Bi atom deposition, 3-AL Bi forms large areas continuous crystalline films. Before the BP substrate is fully covered, 5-AL Bi structures appear when two additional atomic Bi layers start to grown on top of the 3-AL Bi. **b,** The zoomed-in STM image showing the edge at junction of the 3-AL and 5-AL Bi islands. The 5-AL Bi shows a similar STM contrast to that of 3-AL Bi. **c** to **e,** The series STS d*I*/d*V*

spectra measured along the red, blue and pink arrows in (b). In (c), the spectra show an obvious peak belonging to an edge state at -55 mV. When moving slightly away from the edge to the center as shown in (d), the edge state peak decreases in intensity. When far away from the edge in (e), the edge state totally disappears. The inset in (e) shows STS spectra from -150 mV to 300 mV, indicating very similar electronic character of 5-AL Bi to that of the 3-AL Bi. **f,** The d$I$/d$V$ mapping acquired at -55 mV clearly shows the spatial distribution of the edge state of a 5-AL Bi island. The edge of 5-AL Bi is much more ordered than that of 3-AL Bi, most likely because of a negligible hybridization with the BP substrate; this results in a more uniform contrast. The edge state penetrates ~4.0 nm into the island.